\documentclass[aps,prl]{revtex4}
\usepackage{epsfig}

\newcommand{\be}{\begin{equation}}
\newcommand{\ee}{\end{equation}}

\newcommand{\bmath}{\begin{mathletters}}
\newcommand{\emath}{\end{mathletters}}

\begin{document}

\title{\Large{\bf Coexistence of ferromagnetism and superconductivity: the role of kinetic interactions, kinetic correlations, and external pressure}}

\vskip0.5cm 

\author{ Grzegorz G\'{o}rski, Krzysztof Kucab\footnote{Corresponding author. E-mail: kkucab@univ.rzeszow.pl (K.\ Kucab)}, and Jerzy Mizia }

\address{Institute of Physics, Rzesz\'{o}w University, Al. Rejtana 16A, \\
35-959 Rzesz\'{o}w, Poland\\}

\vskip0.5cm 


\begin{abstract}

\noindent
We use the Hubbard type model to describe the coexistence between superconductivity (SC) and
ferromagnetism (F). Our Hamiltonian contains single-site and two-site interactions. All inter-site interactions
will have included the inter-site kinetic correlation: $\langle c_{i\sigma}^{+}c_{j\sigma}\rangle $, within the
Hartree-Fock approximation. To obtain the SC transition temperature $T_{SC}$ and Curie temperature
$T_{C}$ we use the Green's functions method. The numerical results show that the singlet SC is eliminated
by F, but the triplet SC is either enhanced or depleted by F, depending on the carrier concentration and
direction of a superconducting spin pair with respect to magnetization. The kinetic correlation is capable of
creating superconductivity. We find that the ferromagnetism created by change of the bandwidth can coexist
with singlet superconductivity. In the case of triplet superconductivity the ferromagnetism creates different
critical SC temperatures for the $A_{1}$ and $A_{2}$ phase (the pair's spin parallel and antiparallel to
magnetization, respectively).

\end{abstract}
\maketitle


\vskip1.5cm
\noindent {\Large {\bf 1. Introduction}}

\vskip0.5cm 

In the last few years materials have been found where superconductivity (SC) coexists with ferromagnetism (F) within the same electron band. These two cooperative phenomena are mutually antagonistic because superconductivity is associated with the pairing of electron states related to time-reversal, while in the magnetic states the time-reversal symmetry is lost, and therefore there is strong competition between them. However, Ginzburg \cite{Ginzburg1} has pointed out the possibility of this coexistence under the condition that the magnetization is smaller than the thermodynamic critical field multiplied by susceptibility of a given material. Matthias and co-workers \cite{Matthias1} demonstrated that a very small concentration of magnetic impurities was enough to completely destroy superconductivity when ferromagnetic ordering was present.

Superconductivity and ferromagnetism was simultaneously observed in the late 1970s in the intermetallic systems, e.g.\ ErRh$_4$B$_4$ \cite{Sinha1} and HoMo$_6$S$_8$ [4]. These systems have a Curie temperature $T_C$  much lower than the critical superconducting temperature $T_{SC}$. The domain of coexistence is in a very limited narrow temperature range. These compounds also have the second critical temperature $T_{SC2}<T_C$ below which the superconductivity disappears and they become only ferromagnetic. In both: ErRh$_4$B$_4$ and HoMo$_6$S$_8$, the ferromagnetism and superconductivity are created on different atoms, e.g.\ in ErRh$_4$B$_4$ ferromagnetism is carried by 4\textit{f} electrons of Er atoms and the SC by 4\textit{d} electrons of Rh atoms.

There is an intermetallic compound exhibiting a coexistence of SC and F, this is ErNi$_2$B$_2$C \cite{Cava1}, which is characterized by the long range magnetic order associated with magnetic Er ions. This compound has $T_{SC}  \approx 10.5\,{\rm  K}$; below $T_N  \approx 6\,{\rm  K}$ superconductivity coexists with antiferromagnetism, and below $T_{SC}  \approx 2.3\,{\rm  K}$  SC coexists with ferromagnetism \cite{Canfield1}. The magnetic moment of this compound is very weak and is equal to $0.35\,\mu _B/\rm{Er}$ atom at temperature $1.8\,\rm{K}$, which is about $1/22$ of the saturation moment \cite{Kawano1}.

There are also results showing the coexistence between magnetic ordering and superconductivity in a family of hybrid ruthenate-cuprate compounds such as RuSr$_2$GdCu$_2$O$_8$. This compound exhibits a ferromagnetic order at a rather high Curie temperature $T_C  = 133-136\, {\rm K}$, and becomes superconducting at a significantly lower critical temperature $T_{SC}  = 15 - 40\, {\rm K}$ \cite{Bernhard1}.

Recently the so-called ferromagnetic superconductors were discovered, which exhibit simultaneously the ferromagnetic and a spin triplet superconducting phase modified by external pressures. To these ferromagnetic superconductors we include UGe$_2$ \cite{Saxena1,Huxley1}, URhGe \cite{Aoki1}, and ZrZn$_2$ \cite{Pfleiderer1}. Their Curie temperature is much higher than the critical superconducting temperature, and the superconductivity exists only in the presence of ferromagnetism. Both states: ferromagnetism and superconductivity are formed by the same electrons. In UGe$_2$ and URhGe these are 5\textit{f} electrons on U atoms, and in ZrZn$_2$ these are 4\textit{d} electrons of Zr atoms.

The theoretical possibility of ferromagnetism coexisting with the triplet parallel spins superconductivity was suggested by Fay and Appel for ZrZn$_2$ \cite{Fay1}, while the coexistence of F with singlet superconductivity was theoretically developed by Fulde, Ferrell \cite{Fulde1}, and Larkin, Ovchinnikov \cite{Larkin1}. Further theoretical development took place after finding experimental evidence for the coexistence of triplet SC with F (see e.g. \cite{Powell1,Machida1}) or singlet SC with F (see e.g. \cite{Powell1,Cuoco1,Karchev1}).

In this paper we propose the model describing coexistence of ferromagnetism and superconductivity which is based upon the kinetic interactions $\Delta t$ and $t_{ex}$, and the inter-site kinetic correlation $I_\sigma   = \left\langle {c_{i\sigma }^ +  c_{j\sigma } } \right\rangle$. It was shown \cite{Hirsch1,Amadon1,Gorski1} that the ferromagnetic order can be created in two ways: (i) due to mutual shift of majority and minority spin bands, as in the Stoner model, (ii) due to a change in the bandwidth of the spin bands. The first case leads to decrease of potential energy of the system, and the second one leads to lowering of the kinetic energy. The kinetic interactions give rise to both of these mechanisms in creating the ferromagnetic state. They were also used in describing the superconductivity \cite{Hirsch2,Marsiglio1,Micnas1,Micnas2,Arrachea1,Arrachea2,Gorski2}. Hirsch \cite{Hirsch2,Marsiglio1} and Micnas \cite{Micnas2} have shown that the interaction $\Delta t$ leads to s-wave singlet superconductivity in a less then half-filled band system. Aligia and co-workers have demonstrated \cite{Arrachea1,Arrachea2,Aligia1} that the interaction $t_{ex}$ can give rise to both: singlet superconductors with s-wave and d-wave symmetry \cite{Arrachea1,Arrachea2} and triplet superconductors with p-wave symmetry \cite{Aligia1}.

The paper is organized as follows. In Section 2 we have put forward the model Hamiltonian and we analyzed this Hamiltonian using Green's functions technique. In Section 3 we examine the possibility of the coexistence between singlet superconductivity and ferromagnetism. We show the dependence of Curie temperature and superconducting critical temperature on carrier concentration. A similar analysis for triplet superconductivity is presented in Section 4. In Section 5 we present the influence of external pressure on the Curie temperature and the superconducting critical temperature for the triplet superconductors. Section 6 contains conclusions covering results presented in this paper.

\vskip1.0cm 
\noindent {\Large {\bf 2. Model Hamiltonian and Green's functions technique}}

\vskip0.5cm 

The experimental results for ZrZn$_2$ show that the magnetic moment is $0.17\,\mu_B$, which is a small value. This suggests that the ferromagnetic superconductor ZrZn$_2$ is a weak itinerant ferromagnet. It will allow a description of this compound by the extended Stoner model. In this model we include the on-site Coulomb repulsion and the inter-site interactions which modify the electron bandwidth. In the superconducting state we include the experimentally observed spin triplet superconductivity and also the spin singlet superconductivity. Our Hamiltonian can be written in the following form \cite{Gorski3,Mizia1}

\be
H =  - \sum\limits_{ < ij > \atop \scriptstyle \sigma } {\left[ {t - \Delta t\left( {\hat n_{i - \sigma }  + \hat n_{j - \sigma } } \right) + 2t_{ex} \hat n_{i - \sigma } \hat n_{j - \sigma } } \right]c_{i\sigma }^ +  c_{j\sigma }  - \mu _0 \sum\limits_{i\sigma } {\hat n_{i\sigma } } }  + \frac{U}{2}\sum\limits_{i\sigma } {\hat n_{i\sigma } \hat n_{i - \sigma } }  + \frac{J}{2}\sum\limits_{\scriptstyle  < ij >  \hfill \atop 
  \scriptstyle \;\;\sigma \sigma ' \hfill} {c_{i\sigma }^ +  c_{j\sigma '}^ +  c_{i\sigma '} c_{j\sigma } }, 
\label{1}
\ee
where $c_{i\sigma }^ +  \left( {c_{i\sigma } } \right)$ creates (annihilates) electrons with spin $\sigma$ on the lattice site \textit{i}, $\hat n_{i\sigma }  = c_{i\sigma }^ +  c_{i\sigma }$ is the electron number operator for electrons with spin $\sigma$ on the lattice site \textit{i}, \textit{U} is the on-site Coulomb repulsion, \textit{J} is the inter-site exchange interaction, and $\mu_0$ is the chemical potential. Kinetic hopping and exchange hopping interactions: $\Delta t$ and $t_{ex}$, according to \cite{Mizia1}, can be expressed as

\be
\Delta t = t - t_1  = t\left( {1 - S_1 } \right)\;\;\;\; {\rm and} \;\;\;\; t_{ex}  = \frac{{t + t_2 }}{2} - t_1  = t\frac{{1 + S_1 S_2  - 2S_1 }}{2},
\label{2}
\ee
where $S_1$ and $S_2$ are the hopping inhibiting factors

\be
t_1/t=S_1\;\;\; {\rm and}\;\;\; t_2/t_1=S_2,
\label{3}
\ee
$t_1$ and $t_2$ are the electron hopping energies in the presence of one and two electrons with opposite spin, respectively.

After applying the modified Hartree-Fock approximation (H-F), in which the inter-site kinetic correlation $\langle c_{i\sigma}^{+}c_{j\sigma}\rangle $ is included for all inter-site interactions, we obtain the following simplified Hamiltonian (see \S6 of Ref. \cite{Mizia1})

\be
H = H_0  + H_{OSP}  + H_{ESP}. 
\label{4}
\ee
The Hamiltonian $H_0$ is the kinetic energy with added molecular field

\be
H_0  =  - \sum\limits_{ < ij > \atop \scriptstyle \sigma } {t_{eff}^\sigma  c_{i\sigma }^ +  c_{j\sigma } }  - \sum\limits_{i\sigma } {\left( {\mu _0  - M_\sigma  } \right)\hat n_{i\sigma } }, 
\label{5}
\ee
where the effective hopping interaction is $t_{eff}^\sigma   = tb^\sigma $, the bandwidth modification factor has the form

\be
b^\sigma   = 1 - \frac{1}{t}\left[ {2\Delta tn_{ - \sigma }  - 2t_{ex} \left( {n_{ - \sigma }^2  - I_{ - \sigma }^2  - 2I_\sigma  I_{ - \sigma }  - \left| {\Delta _0 } \right|^2  - \left| {\Delta _h^S } \right|^2  + \left| {\Delta _h^\sigma  } \right|^2 } \right) + J\left( {I_\sigma   + I_{ - \sigma } } \right)} \right],
\label{6}
\ee
and the spin-dependent modified molecular field is given by
\be
M_\sigma   = Un_{ - \sigma }  - zJn_\sigma   + 2z\Delta tI_\sigma   - 2zt_{ex} \left( 2I_{ - \sigma } n_\sigma   + \Delta _h^S \Delta _0^* + \Delta _h^{S*}  \Delta _0  \right ),
\label{7}
\ee
where \textit{z} is the number of nearest neighbors.

The electron occupation number used above is $n_\sigma   = \left\langle {c_{i\sigma }^ +  c_{i\sigma } } \right\rangle$ and the Fock's parameter, proportional to the kinetic energy, is given by $I_\sigma   = \left\langle {c_{i\sigma }^ +  c_{j\sigma } } \right\rangle$. The bandwidth change factor $b^\sigma$ includes the following averages

\be
\Delta _0  = \frac{1}{2}\sum\limits_\sigma  {\sigma \left\langle {c_{i - \sigma } c_{i\sigma } } \right\rangle }, \;\;\; \Delta _h^S  = \frac{1}{2}\sum\limits_\sigma  {\sigma \left\langle {c_{i + h, - \sigma } c_{i\sigma } } \right\rangle }, \;\;\; \Delta _h^\sigma   = \left\langle {c_{i + h,\sigma } c_{i\sigma } } \right\rangle,
\label{8}
\ee
where $h=j-i$ is the difference of lattice indices for the nearest neighboring atoms in a given direction, $\Delta_0$ and $\Delta_h^S$ are the single-site and inter-site singlet superconducting parameters, and $\Delta_h^\sigma$ is the inter-site equal spin triplet superconducting parameter.

The next two terms of the Hamiltonian (\ref{4}), $H_{OSP}$ and $H_{ESP}$, are related to the opposite spin pairing (OSP) and equal (parallel) spin pairing (ESP), respectively. The OSP term describes the singlet SC with the total spin being $0$ and the triplet SC with total spin $1$ (in units of $\hbar$) and its projection being $0$ (see \cite{Leggett1}). This term is given by

\be
H_{OSP}  = \sum\limits_i {a_1 \left( {c_{i \uparrow }^ +  c_{i \downarrow }^ +   + h.c.} \right)}  + \sum\limits_{ < ij > } {a_2 \left( {c_{i \uparrow }^ +  c_{j \downarrow }^ +   - c_{i \downarrow }^ +  c_{j \uparrow }^ +   + h.c.} \right)}  + \sum\limits_{ < ij > } {a_3 \left( {c_{i \uparrow }^ +  c_{j \downarrow }^ +   + c_{i \downarrow }^ +  c_{j \uparrow }^ +   + h.c.} \right)}, 
\label{9}
\ee
where

\be
a_1  = \left[ {U + 2zt_{ex} \left( {I_ \uparrow   + I_ \downarrow  } \right)} \right]\Delta _0  + 2z\left( {\Delta t - t_{ex} n} \right)\Delta _h^S, 
\label{10}
\ee

\be
a_2  = \left( {\Delta t - t_{ex} n} \right)\Delta _0  + \frac{1}{2}\left[ {J + 2t_{ex} \left( {I_ \uparrow   + I_ \downarrow  } \right)} \right]\Delta _h^S, 
\label{11}
\ee

\be
a_3  = \frac{1}{2}\left[ { - J + 2t_{ex} \left( {I_ \uparrow   + I_ \downarrow  } \right)} \right]\Delta _h^T, 
\label{12}
\ee
and

\be
\Delta _h^T  = \frac{1}{2}\sum\limits_\sigma  {\left\langle {c_{i + h, - \sigma } c_{i\sigma } } \right\rangle } 
\label{13}
\ee
is the inter-site opposite spin triplet superconducting parameter.

\noindent
The ESP term describes the triplet SC with total spin $1$, its projection being $\pm 1$ \cite{Leggett1}, and is given by

\be
H_{ESP}  = \frac{1}{2}\sum\limits_{ < ij > \atop \scriptstyle \sigma } {\left( { - J - 4t_{ex} I_{ - \sigma } } \right)\Delta _h^\sigma  \left( {c_{i\sigma }^ +  c_{j\sigma }^ +   + h.c.} \right)}. 
\label{14}
\ee
Transforming Hamiltonian (\ref{4}) into the momentum space we obtain

\be
H = \sum\limits_{k\sigma } {\varepsilon _k^\sigma  \hat n_{k\sigma } }  - \sum\limits_k {\left[ {\left( {\Delta _{kS}^{ \uparrow  \downarrow }  + \Delta _{kT}^{ \uparrow  \downarrow } } \right)c_{k \uparrow }^ +  c_{ - k \downarrow }^ +   + h.c.} \right]}  - \sum\limits_{k\sigma } {\left( {\Delta _k^\sigma  c_{k\sigma }^ +  c_{ - k\sigma }^ +   + h.c.} \right)}, 
\label{15}
\ee
where

\be
\varepsilon _k^\sigma   = \varepsilon _k b^\sigma   + M_\sigma   - \mu _0 
\label{16}
\ee
is the spin dependent modified dispersion relation, $\varepsilon_k$ is the unperturbed dispersion relation which has the following form

\be
\varepsilon _k  =  - t\sum\limits_h {e^{i{\bf k} \cdot {\bf h}} }, 
\label{17}
\ee
where ${\bf h} = {\bf r}_i  - {\bf r}_j$ is pointing to the nearest neighbors lattice sites.

The singlet superconductivity energy gap $\Delta _{kS}^{ \uparrow  \downarrow }$ is the Fourier transformation of the first two terms in the Hamiltonian (\ref{9}). It is symmetric, $\Delta _{ - kS}^{ \uparrow  \downarrow }  = \Delta _{kS}^{ \uparrow  \downarrow }$, and is given by the relation

\be
\Delta _{kS}^{ \uparrow  \downarrow }  = d_0  + d_1 \Delta _{kS}^{\left( 1 \right)}  + d_2 \Delta _{kS}^{\left( 2 \right)}, 
\label{18}
\ee
where

\be
\begin{array}{l}
d_0  =  - \left[ {U + 2zt_{ex} \left( {I_ \uparrow   + I_ \downarrow  } \right)} \right]\Delta _0  - 2z\left( {\Delta t - t_{ex} n} \right)\Delta _h^S, \\
d_1  =  - \left( {\Delta t - t_{ex} n} \right),\\
d_2  =  - \frac{1}{2}\left[ {J + 2t_{ex} \left( {I_ \uparrow   + I_ \downarrow  } \right)} \right].\\
\end{array}
\label{19}
\ee
The momentum independent gap parameter above is given by a Fourier transformation of Eq.\ (\ref{8}) to momentum space

\be
\Delta _0  = \frac{1}{{2N}}\sum\limits_{k\sigma } {\sigma \left\langle {c_{ - k - \sigma } c_{k\sigma } } \right\rangle },
\label{20}
\ee
and the momentum dependent gap parameters are given as

\be
\Delta _{kS}^{\left( 1 \right)}  = \Delta _0 \sum\limits_h {\left( {e^{i{\bf k} \cdot {\bf h}}  + e^{ - i{\bf k} \cdot {\bf h}} } \right)}\;,\;\;\; \Delta _{kS}^{\left( 2 \right)}  = \sum\limits_h {\Delta _h^S \left( {e^{i{\bf k} \cdot {\bf h}}  + e^{ - i{\bf k} \cdot {\bf h}} } \right)}.  
\label{21}
\ee

The opposite spin triplet energy gap $\Delta _{kT}^{ \uparrow  \downarrow }$ is the Fourier transformation of the third term in the Hamiltonian (\ref{9}), it is antisymmetrical, $\Delta _{ - kT}^{ \uparrow  \downarrow }  =  - \Delta _{kT}^{ \uparrow  \downarrow }$, and is given by

\be
\Delta _{kT}^{ \uparrow  \downarrow }  = d_3 \Delta _{kT}, 
\label{22}
\ee
where

\be
d_3  = \frac{1}{2}\left[ {J - 2t_{ex} \left( {I_ \uparrow   + I_ \downarrow  } \right)} \right],
\label{23}
\ee
and

\be
\Delta _{kT}  = \sum\limits_h {\Delta _h^T \left( {e^{i{\bf k} \cdot {\bf h}}  - e^{ - i{\bf k} \cdot {\bf h}} } \right)}. 
\label{24}
\ee

The triplet superconducting ordering parameter for parallel spins is denoted by $\Delta_k^\sigma$ and it is the Fourier transformation of $\Delta_h^\sigma$ from Eq.\ (\ref{8}). It is given by the relation

\be
\Delta _k^\sigma   = d_4 \Delta _{kT}^\sigma,
\label{25}
\ee
where

\be
d_4  = \frac{1}{2}\left( {J + 4t_{ex} I_{ - \sigma } } \right),
\label{26}
\ee
and

\be
\Delta _{kT}^\sigma   = \sum\limits_h {\Delta _h^\sigma  e^{i{\bf k} \cdot {\bf h}} }.
\label{27}
\ee

The Hamiltonian (\ref{15}) will be analyzed using Green's functions technique. The Green's function $\langle \langle A;B\rangle \rangle _\varepsilon$ satisfies the following equation of motion

\be
\varepsilon \langle \langle A;B\rangle \rangle _\varepsilon   = \left\langle {\left[ {A,B} \right]_ +  } \right\rangle  + \langle \langle \left[ {A,H} \right]_ -  ;B\rangle \rangle _\varepsilon.
\label{28}
\ee
Using Hamiltonian (\ref{15}) in the equation above we obtain a set of equations, which can be written as

\be
\left( {\begin{array}{*{20}c}
   {\varepsilon  - \varepsilon _k^ \uparrow  } & 0 & {2\Delta _k^ \uparrow  } & {\Delta _{kS}^{ \uparrow  \downarrow }  + \Delta _{kT}^{ \uparrow  \downarrow } }  \\
   0 & {\varepsilon  - \varepsilon _k^ \downarrow  } & { - \Delta _{kS}^{ \uparrow  \downarrow }  + \Delta _{kT}^{ \uparrow  \downarrow } } & {2\Delta _k^ \downarrow  }  \\
   { - 2(\Delta _{ - k}^\uparrow )^* } & { - (\Delta _{ - kS}^{ \uparrow  \downarrow })^*  - (\Delta _{ - kT}^{ \uparrow  \downarrow })^* } & {\varepsilon  + \varepsilon _{ - k}^ \uparrow  } & 0  \\
   {(\Delta _{ - kS}^{ \uparrow  \downarrow })^*  - (\Delta _{ - kT}^{ \uparrow  \downarrow })^* } & { - 2(\Delta _{ - k}^ {\downarrow  })^* } & 0 & {\varepsilon  + \varepsilon _{ - k}^ \downarrow  }  \\
\end{array}} \right){\bf \hat G}\left( {k,\varepsilon } \right) = {\bf \hat 1},
\label{29}
\ee
where ${\bf \hat 1}$ is the unity matrix and the Green's functions matrix has the form

\be
{\bf \hat G}\left( {k,\varepsilon } \right) = \left( {\begin{array}{*{20}c}
   {\langle \langle c_{k \uparrow } ;c_{k \uparrow }^ +  \rangle \rangle _\varepsilon  } & {\langle \langle c_{k \uparrow } ;c_{k \downarrow }^ +  \rangle \rangle _\varepsilon  } & {\langle \langle c_{k \uparrow } ;c_{ - k \uparrow } \rangle \rangle _\varepsilon  } & {\langle \langle c_{k \uparrow } ;c_{ - k \downarrow } \rangle \rangle _\varepsilon  }  \\
   {\langle \langle c_{k \downarrow } ;c_{k \uparrow }^ +  \rangle \rangle _\varepsilon  } & {\langle \langle c_{k \downarrow } ;c_{k \downarrow }^ +  \rangle \rangle _\varepsilon  } & {\langle \langle c_{k \downarrow } ;c_{ - k \uparrow } \rangle \rangle _\varepsilon  } & {\langle \langle c_{k \downarrow } ;c_{ - k \downarrow } \rangle \rangle _\varepsilon  }  \\
   {\langle \langle c_{ - k \uparrow }^ +  ;c_{k \uparrow }^ +  \rangle \rangle _\varepsilon  } & {\langle \langle c_{ - k \uparrow }^ +  ;c_{k \downarrow }^ +  \rangle \rangle _\varepsilon  } & {\langle \langle c_{ - k \uparrow }^ +  ;c_{ - k \uparrow } \rangle \rangle _\varepsilon  } & {\langle \langle c_{ - k \uparrow }^ +  ;c_{ - k \downarrow } \rangle \rangle _\varepsilon  }  \\
   {\langle \langle c_{ - k \downarrow }^ +  ;c_{k \uparrow }^ +  \rangle \rangle _\varepsilon  } & {\langle \langle c_{ - k \downarrow }^ +  ;c_{k \downarrow }^ +  \rangle \rangle _\varepsilon  } & {\langle \langle c_{ - k \downarrow }^ +  ;c_{ - k \uparrow } \rangle \rangle _\varepsilon  } & {\langle \langle c_{ - k \downarrow }^ +  ;c_{ - k \downarrow } \rangle \rangle _\varepsilon  }  \\
\end{array}} \right).
\label{30}
\ee
Using the symmetry properties of the order parameter we can rewrite Eq.\ (\ref{29}) in the following form

\be
\left( {\begin{array}{*{20}c}
   {\varepsilon  - \varepsilon _k^ \uparrow  } & 0 & {2\Delta _k^ \uparrow  } & {\Delta _{kS}^{ \uparrow  \downarrow }  + \Delta _{kT}^{ \uparrow  \downarrow } }  \\
   0 & {\varepsilon  - \varepsilon _k^ \downarrow  } & { - \Delta _{kS}^{ \uparrow  \downarrow }  + \Delta _{kT}^{ \uparrow  \downarrow } } & {2\Delta _k^ \downarrow  }  \\
   {2(\Delta _k^ {\uparrow  })^* } & { - (\Delta _{kS}^{ \uparrow  \downarrow })^*  + (\Delta _{kT}^{ \uparrow  \downarrow })^* } & {\varepsilon  + \varepsilon _{ - k}^ \uparrow  } & 0  \\
   {(\Delta _{kS}^{ \uparrow  \downarrow })^*  + (\Delta _{kT}^{ \uparrow  \downarrow })^* } & {2(\Delta _k^{ \downarrow  })^* } & 0 & {\varepsilon  + \varepsilon _{ - k}^ \downarrow  }  \\
\end{array}} \right){\bf \hat G}\left( {k,\varepsilon } \right) = {\bf \hat 1}.
\label{31}
\ee
In further analysis we will consider separately, the coexistence of ferromagnetism with the singlet, and with the ESP triplet superconductivity.

\vskip1.0cm 
\noindent {\Large {\bf 3. Coexistence of ferromagnetism and singlet superconductivity}}

\vskip0.5cm 

In this case the energy gaps in Eq.\ (\ref{31}) are: $\Delta_k^\sigma = 0$, $\Delta _{kT}^{ \uparrow  \downarrow }  = 0$, and $\Delta _{kS}^{ \uparrow  \downarrow }  \ne 0$ is given by Eq.\ (\ref{18}). This assumptions simplifies the Green's functions ${\bf \hat G}\left( {k,\varepsilon } \right)$. Using in Eq.\ (\ref{8}) the Zubarev's relation \cite{Zubarev1,Hubbard1} for the average of the operators' product, and the Green's functions ${\bf \hat G}\left( {k,\varepsilon } \right)$ from Eq.\ (\ref{31}), we calculate the on-site superconductivity parameter $\Delta_0$

\be
\Delta _0  = \frac{1}{2}\sum\limits_\sigma  {\sigma \left\langle {c_{i - \sigma } c_{i\sigma } } \right\rangle }  = \frac{1}{{2N}}\sum\limits_{k\sigma } {\sigma \left\langle {c_{ - k - \sigma } c_{k\sigma } } \right\rangle }  =  - \frac{1}{{2N}}\sum\limits_k {\frac{1}{\pi }\int {f\left( \varepsilon  \right){\mathop{\rm Im}\nolimits} } \left( {\langle \langle c_{k \uparrow } ;c_{ - k \downarrow } \rangle \rangle _\varepsilon   - \langle \langle c_{k \downarrow } ;c_{ - k \uparrow } \rangle \rangle _\varepsilon  } \right)d\varepsilon }, 
\label{32}
\ee 
and the inter-site superconductivity parameter $\Delta_h^S$

\be
\Delta _h^S  = \frac{1}{2}\sum\limits_\sigma  {\sigma \left\langle {c_{i + h, - \sigma } c_{i\sigma } } \right\rangle }  = \frac{1}{{2N}}\sum\limits_{k\sigma } {e^{i{\bf k} \cdot {\bf h}} \sigma \left\langle {c_{ - k - \sigma } c_{k\sigma } } \right\rangle }  =  - \frac{1}{{2N}}\sum\limits_k {e^{i{\bf k} \cdot {\bf h}} \frac{1}{\pi }\int {f\left( \varepsilon  \right){\mathop{\rm Im}\nolimits} } \left( {\langle \langle c_{k \uparrow } ;c_{ - k \downarrow } \rangle \rangle _\varepsilon   - \langle \langle c_{k \downarrow } ;c_{ - k \uparrow } \rangle \rangle _\varepsilon  } \right)d\varepsilon }. 
\label{33}
\ee
To solve the above equations we obtain functions $\langle \langle c_{k \uparrow } ;c_{ - k \downarrow } \rangle \rangle _\varepsilon$ and $\langle \langle c_{k \downarrow } ;c_{ - k \uparrow } \rangle \rangle _\varepsilon$ from Eq.\ (\ref{31}). They have the following forms

\be
\langle \langle c_{k \uparrow } ;c_{ - k \downarrow } \rangle \rangle _\varepsilon   =  - \frac{{\Delta _{kS}^{ \uparrow  \downarrow } }}{{\left( {\varepsilon  - \varepsilon _k^ \uparrow  } \right)\left( {\varepsilon  + \varepsilon _{ - k}^ \downarrow  } \right) - \left( {\Delta _{kS}^{ \uparrow  \downarrow } } \right)^2 }},
\label{34}
\ee
and

\be
\langle \langle c_{k \downarrow } ;c_{ - k \uparrow } \rangle \rangle _\varepsilon   = \frac{{\Delta _{kS}^{ \uparrow  \downarrow } }}{{\left( {\varepsilon  - \varepsilon _k^ \downarrow  } \right)\left( {\varepsilon  + \varepsilon _{ - k}^ \uparrow  } \right) - \left( {\Delta _{kS}^{ \uparrow  \downarrow } } \right)^2 }}.
\label{35}
\ee
Inserting these functions to Eqs.\ (\ref{32}) and (\ref{33}) we obtain

\be
\Delta _0  =  - \frac{1}{{2N}}\sum\limits_k {\Delta _{kS}^{ \uparrow  \downarrow } \frac{1}{\pi }\int {f\left( \varepsilon  \right){\mathop{\rm Im}\nolimits} \left[ {G\left( {k,\varepsilon  - \varepsilon _{k1} } \right)G\left( {k, - \varepsilon  + \varepsilon _{k1} } \right) + G\left( {k,\varepsilon  + \varepsilon _{k1} } \right)G\left( {k, - \varepsilon  - \varepsilon _{k1} } \right)} \right]d\varepsilon } }, 
\label{36}
\ee
and

\be
\Delta _h^S  =  - \frac{1}{{2N}}\sum\limits_k {e^{i{\bf k} \cdot {\bf h}} \Delta _{kS}^{ \uparrow  \downarrow } \frac{1}{\pi }\int {f\left( \varepsilon  \right){\mathop{\rm Im}\nolimits} \left[ {G\left( {k,\varepsilon  - \varepsilon _{k1} } \right)G\left( {k, - \varepsilon  + \varepsilon _{k1} } \right) + G\left( {k,\varepsilon  + \varepsilon _{k1} } \right)G\left( {k, - \varepsilon  - \varepsilon _{k1} } \right)} \right]d\varepsilon } }, 
\label{37}
\ee
where

\be
G\left( {k,\varepsilon } \right) = \frac{1}{{\varepsilon  - E_k }},\;\;\;E_k  = \sqrt {\varepsilon _{k0}^2  + \left( {\Delta _{kS}^{ \uparrow  \downarrow } } \right)^2 },
\label{38}
\ee

\be
\varepsilon _{k0}  = \frac{{\varepsilon _k^ \uparrow   + \varepsilon _k^ \downarrow  }}{2},\;\;\;\varepsilon _{k1}  = \frac{{\varepsilon _k^ \uparrow   - \varepsilon _k^ \downarrow  }}{2},
\label{39}
\ee
and $\Delta _{kS}^{ \uparrow  \downarrow }$ is given by relations (\ref{18}) and (\ref{19}). The Fock's parameter appearing in Eqs.\ (\ref{19}) is the Fourier transformation of $I_\sigma   = \left\langle {c_{i\sigma }^ +  c_{j\sigma } } \right\rangle$. Based on the Zubarev's relation it can be calculated as

\be
I_\sigma   =  - \frac{1}{{2N}}\sum\limits_k {\left( {\frac{1}{z}\sum\limits_h {e^{i{\bf k} \cdot {\bf h}} } } \right)\frac{1}{\pi }\int {f\left( \varepsilon  \right){\mathop{\rm Im}\nolimits} \left( {\langle \langle c_{k \uparrow } ;c_{k \uparrow }^ +  \rangle \rangle _\varepsilon   - \langle \langle c_{ - k \uparrow }^ +  ;c_{ - k \uparrow } \rangle \rangle _\varepsilon  } \right)d\varepsilon } }.
\label{40}
\ee
The above equations together with the equations for electron concentration and magnetization

\be
n = 1 + \frac{1}{{2N}}\sum\limits_k {\frac{1}{\pi }\int {f\left( \varepsilon  \right){\mathop{\rm Im}\nolimits} \left( {\langle \langle c_{k \uparrow } ;c_{k \uparrow }^ +  \rangle \rangle _\varepsilon   + \langle \langle c_{k \downarrow } ;c_{k \downarrow }^ +  \rangle \rangle _\varepsilon   - \langle \langle c_{ - k \uparrow }^ +  ;c_{ - k \uparrow } \rangle \rangle _\varepsilon   - \langle \langle c_{ - k \downarrow }^ +  ;c_{ - k \downarrow } \rangle \rangle _\varepsilon  } \right)d\varepsilon } },
\label{41}
\ee

\be
m = \frac{1}{N}\sum\limits_k {\frac{1}{\pi }\int {f\left( \varepsilon  \right){\mathop{\rm Im}\nolimits} \left( {\langle \langle c_{k \uparrow } ;c_{k \uparrow }^ +  \rangle \rangle _\varepsilon   - \langle \langle c_{k \downarrow } ;c_{k \downarrow }^ +  \rangle \rangle _\varepsilon   - \langle \langle c_{ - k \uparrow }^ +  ;c_{ - k \uparrow } \rangle \rangle _\varepsilon   + \langle \langle c_{ - k \downarrow }^ +  ;c_{ - k \downarrow } \rangle \rangle _\varepsilon  } \right)d\varepsilon } },
\label{42}
\ee
constitute the set of self-consistent equations for parameters of ferromagnetic and superconducting states.

The parameters $\Delta_0$ and $\Delta_h^S$, described by Eqs.\ (\ref{36}) and (\ref{37}), can be calculated by the help of identity

\be
\frac{1}{{\varepsilon  - \varepsilon _k  + i0^ +  }} = P\left( {\frac{1}{{\varepsilon  - \varepsilon _k }}} \right) - i\pi \delta \left( {\varepsilon  - \varepsilon _k } \right),
\label{43}
\ee
where \textit{P} is the principal value of the integral and the symbol $0^+$ stands for the infinitesimal positive value. As a result we obtain

\be
\Delta _0  = \frac{1}{{4N}}\sum\limits_k {\frac{{\Delta _{kS}^{ \uparrow  \downarrow } }}{{E_k }}\left( {\tanh \frac{{E_k  + \varepsilon _{k1} }}{{2k_B T}} + \tanh \frac{{E_k  - \varepsilon _{k1} }}{{2k_B T}}} \right)},
\label{44}
\ee

\be
\Delta _h^S  = \frac{1}{{4N}}\sum\limits_k {e^{i{\bf k} \cdot {\bf h}} \frac{{\Delta _{kS}^{ \uparrow  \downarrow } }}{{E_k }}\left( {\tanh \frac{{E_k  + \varepsilon _{k1} }}{{2k_B T}} + \tanh \frac{{E_k  - \varepsilon _{k1} }}{{2k_B T}}} \right)}.
\label{45}
\ee
In a similar way we obtain from the Eqs.\ (\ref{41}), (\ref{42}), and (\ref{40}) the following relations

\be
n = 1 - \frac{1}{{2N}}\sum\limits_k {\left[ {\frac{{\varepsilon _{k0} }}{{E_k }}\left( {\tanh \frac{{E_k  + \varepsilon _{k1} }}{{2k_B T}} + \tanh \frac{{E_k  - \varepsilon _{k1} }}{{2k_B T}}} \right)} \right]},
\label{46}
\ee

\be
m = \frac{1}{{2N}}\sum\limits_k {\left( {\tanh \frac{{E_k  + \varepsilon _{k1} }}{{2k_B T}} - \tanh \frac{{E_k  - \varepsilon _{k1} }}{{2k_B T}}} \right)},
\label{47}
\ee

\be
I_\sigma   =  - \frac{1}{{2N}}\sum\limits_k {\left( {\frac{1}{z}\sum\limits_h {e^{i{\bf k} \cdot {\bf h}} } } \right)\left( {\frac{{E_k  + \varepsilon _{k0} }}{{2E_k }}\tanh \frac{{E_k  + \sigma \varepsilon _{k1} }}{{2k_B T}} - \frac{{E_k  - \varepsilon _{k0} }}{{2E_k }}\tanh \frac{{E_k  - \sigma \varepsilon _{k1} }}{{2k_B T}}} \right)}.
\label{48}
\ee
Inserting Eqs.\ (\ref{44}) and (\ref{45}) into Eqs.\ (\ref{18}) and (\ref{19}), and using the above relations for carrier concentration, magnetization, and Fock's parameter, we can analyze the critical temperature versus carrier concentration for the singlet superconductors. The results are shown in Fig.\ 1. For simplicity it was assumed that

\be
I_\sigma   = n_\sigma  \left( {1 - n_\sigma  } \right),
\label{49}
\ee
instead of Eq.\ (\ref{48}). This result is strict for the constant density of states (DOS) at zero temperature.

\begin{figure}[t]
\begin{center}
\epsfig{file=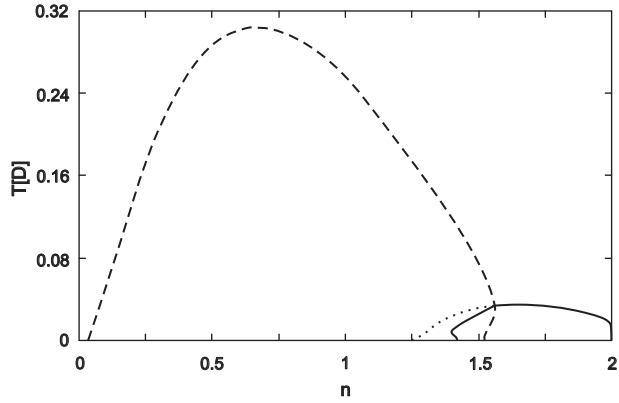,width=0.45\hsize}
	\end{center}
	\caption {Dependence of critical superconducting temperature for the singlet superconductivity (solid line) and Curie temperature (dashed line) on carrier concentration \textit{n}. Dotted line shows the critical SC temperature vs. carrier concentration without ferromagnetism. Calculations were performed for $J=0.0875D$, $U=0.75D$, and $D=0.75\, {\rm eV}$. The hopping inhibiting factors are: $S_1=0.5$ and $S_2=0.05$.}
\vskip0.5 cm
\end{figure}

The critical temperatures shown in Fig.\ 1 depend on all interactions appearing in the Hamiltonian (\ref{1}). In these and future calculations we use the dispersion relation for the 2 dimensional simple cubic lattice.

As can be seen from this figure, the superconductivity created by kinetic interactions is not entirely pushed away by ferromagnetic ordering. In the interval of carrier concentrations around and below $1.5$ superconductivity and ferromagnetism coexists together. The main driving force for the singlet superconductivity is the hopping interaction $\Delta t$. This interaction creates superconductivity by contributing to the negative pairing potential and also by decreasing the width of both spin sub-bands. The decrease of the majority spin sub-band width causes a gain in kinetic energy, which in turn creates a ferromagnetic state \cite{Gorski1}. Such a ferromagnetic state, created by bandwidth changes, can coexist with the superconductivity \cite{Cuoco1}. This case is opposite to the effect of the band shift (i.e.\ Stoner model), which creates ferromagnetism but destroys superconductivity.

\vskip1.0cm 
\noindent {\Large {\bf 4. Coexistence of ferromagnetism and triplet equal spin pairing (ESP) superconductivity}}

\vskip0.5cm 

The experimental evidence shows \cite{Santi1,Shen1} that only the triplet parallel spins superconductivity can coexist with ferromagnetism within the same band of electrons. To explain this effect we assume in Eq.\ (\ref{31}) that the opposite spins parameters are zero ($\Delta _{kS}^{ \uparrow  \downarrow }  = 0$ and $\Delta _{kT}^{ \uparrow  \downarrow }  = 0$) and we obtain the following relation

\be
\left( {\begin{array}{*{20}c}
   {\varepsilon  - \varepsilon _k^ \uparrow  } & 0 & {2\Delta _k^ \uparrow  } & 0  \\
   0 & {\varepsilon  - \varepsilon _k^ \downarrow  } & 0 & {2\Delta _k^ \downarrow  }  \\
   {2(\Delta _k^ {\uparrow  })^* } & 0 & {\varepsilon  + \varepsilon _{ - k}^ \uparrow  } & 0  \\
   0 & {2(\Delta _k^ {\downarrow  })^* } & 0 & {\varepsilon  + \varepsilon _{ - k}^ \downarrow  }  \\
\end{array}} \right){\bf \hat G}\left( {k,\varepsilon } \right) = {\bf \hat 1},
\label{50}
\ee
where the energy gap $\Delta_k^\sigma$ is given by Eq.\ (\ref{25}).

\noindent
Solving Eqs.\ (\ref{50}) and (\ref{25}) in a similar way to the previous case, we obtain the following equations for the superconductivity

\be
\Delta _h^\sigma   =  - \frac{1}{N}\sum\limits_k {e^{i{\bf k} \cdot {\bf h}} \Delta _k^\sigma  \frac{1}{\pi }\int {f\left( \varepsilon  \right){\mathop{\rm Im}\nolimits} \left[ {G^\sigma  \left( {k,\varepsilon } \right)G^\sigma  \left( {k, - \varepsilon } \right)} \right]d\varepsilon } }, 
\label{51}
\ee
where

\be
G^\sigma  \left( {k,\varepsilon } \right) = \left( {\varepsilon  - E_k^\sigma  } \right)^{ - 1},\;\;\;E_k^\sigma   = \sqrt {\left( {\varepsilon _k^\sigma  } \right)^2  + \left( {2\Delta _k^\sigma  } \right)^2 }.
\label{52}
\ee
In the H-F approximation, the superconductivity equation takes on the following form

\be
\Delta _h^\sigma   = \frac{1}{N}\sum\limits_k {e^{i{\bf k} \cdot {\bf h}} \frac{{\Delta _k^\sigma  }}{{E_k^\sigma  }}\tanh \frac{{E_k^\sigma  }}{{2k_B T}}}. 
\label{53}
\ee
From expressions (\ref{41}), (\ref{42}), and (\ref{40}) the equations for carrier concentration, magnetization, and Fock's parameter can be calculated as

\be
n = 1 - \frac{1}{N}\sum\limits_k {\left( {\frac{{\varepsilon _k^ \uparrow  }}{{2E_k^ \uparrow  }}\tanh \frac{{E_k^ \uparrow  }}{{2k_B T}} + \frac{{\varepsilon _k^ \downarrow  }}{{2E_k^ \downarrow  }}\tanh \frac{{E_k^ \downarrow  }}{{2k_B T}}} \right)},
\label{54}
\ee

\be
m = \frac{1}{N}\sum\limits_k {\left( {\frac{{\varepsilon _k^ \downarrow  }}{{2E_k^ \downarrow  }}\tanh \frac{{E_k^ \downarrow  }}{{2k_B T}} - \frac{{\varepsilon _k^ \uparrow  }}{{2E_k^ \uparrow  }}\tanh \frac{{E_k^ \uparrow  }}{{2k_B T}}} \right)}, 
\label{55}
\ee

\be
I_\sigma   = \frac{1}{{2N}}\sum\limits_k {\left( {\frac{1}{z}\sum\limits_h {e^{i{\bf k} \cdot {\bf h}} } } \right)\left( {1 - \frac{{\varepsilon _k^\sigma  }}{{E_k^\sigma  }}\tanh \frac{{E_k^\sigma  }}{{2k_B T}}} \right)}. 
\label{56}
\ee

Equations (\ref{53})--(\ref{56}) are a set of self-consistent complex equations describing the ESP superconductivity coexisting with ferromagnetism. From this set we can find the dependence of the critical temperatures on carrier concentration. The results are shown in Fig.\ 2 for the A$_1$ phase (where only parallel to magnetization pairing occurs) and for the A$_2$ phase (where antiparallel to magnetization pairing exists).

\begin{figure}[t]
\begin{center}
\epsfig{file=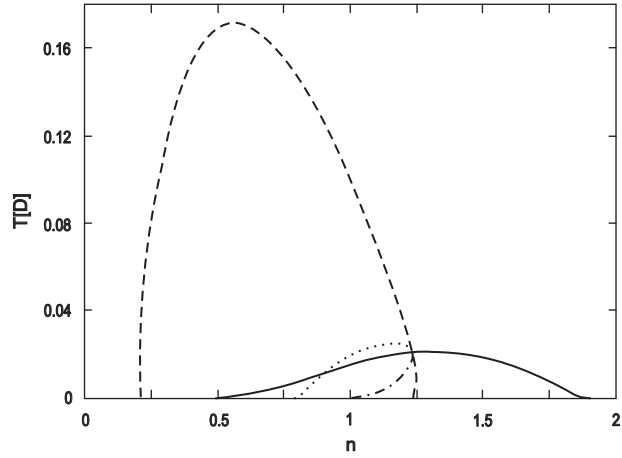,width=0.45\hsize}
	\end{center}
	\caption{Dependence of the superconducting critical temperature $T_{SC}$ and ferromagnetic Curie temperature $T_C$ on carrier concentration $n$ in the case of ESP for $J=0.0875D$, $S_1=S_2=0.4$, $U=0$, and $D=0.5\, {\rm eV}$. Dashed line -- Curie temperature; solid line -- superconducting critical temperature without ferromagnetism; dot-dashed line -- $T_{SC}$ for A$_1$ phase; dotted line --   $T_{SC}$ for A$_2$ phase.}
\vskip0.5 cm
\end{figure}

For the superconducting effect without ferromagnetism the dependence of critical superconducting temperature on carrier concentration is determined by two factors: pairing potential $d_4$ given by Eq.\ (\ref{26}) and the change in kinetic energy described by the bandwidth modification factor $b^\sigma  \left( n \right)$. Pairing potential $d_4$ is symmetric with respect to $n=1$, where it reaches its maximum value. For the parameters $J=0.0875D$, $S_1=S_2=0.4$, and $D=0.5\,{\rm eV}$  we obtain ${{\partial b^\sigma  } \mathord{\left/ {\vphantom {{\partial b^\sigma  } {\partial n}}} \right. \kern-\nulldelimiterspace} {\partial n}} < 0$ at $n<1.6$ and ${{\partial b^\sigma  } \mathord{\left/ {\vphantom {{\partial b^\sigma  } {\partial n}}} \right. \kern-\nulldelimiterspace} {\partial n}} > 0$ at $n>1.6$. The decrease of $b^\sigma$ factor causes an increase of the critical temperature. The net result of both these effects gives the maximum of $T_{SC}(n)$ around $n\approx 1.25$. When the inter-site correlations are not taken into account, $b^\sigma(n)=const$, then the dependence $T_{SC}(n)$ is determined solely by pairing potential. As a result the maximum of the critical temperature is localized at $n=1$ (see e.g.\ \cite{Micnas1}).

At concentrations with nonzero magnetization: $n<1.23$, we have different $T_{SC}$ for A$_1$ and A$_2$ phases. This difference in $T_{SC}(n)$ is determined mainly by ${{\partial b^\sigma  } \mathord{\left/ {\vphantom {{\partial b^\sigma  } {\partial m}}} \right. \kern-\nulldelimiterspace} {\partial m}}$. For A$_2$ phase the increase in magnetization causes a decrease of the bandwidth modification factor: $\partial b^\downarrow/\partial m <0$ (see Fig.\ 3), that in effect increases the critical temperature. For A$_1$ phase we have: $\partial b^\uparrow/\partial m >0$, and the decrease in $T_{SC}$. The second factor favoring A$_2$ phase at $n>1$ is the shift of Fermi level for the minority carriers towards the center of the band, where there are higher critical temperatures \cite{Micnas1}.

\begin{figure}[t]
\begin{center}
\epsfig{file=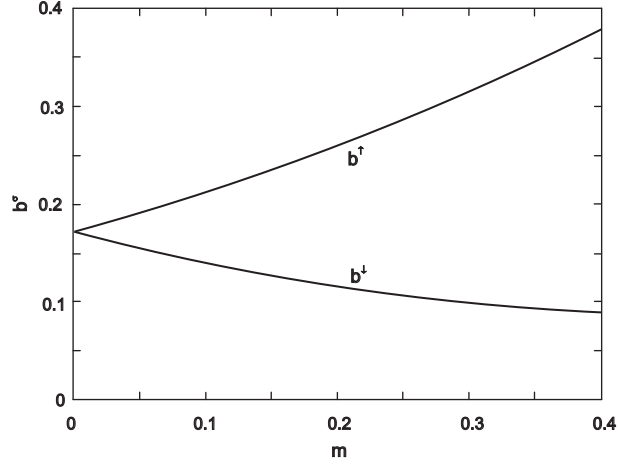,width=0.45\hsize}
	\end{center}
	\caption{Dependence of the bandwidth modification factor $b^\sigma$ on magnetization $m$ for $n=1.23$, $J=0.0875D$, and $S_1=S_2=0.4$.}
\vskip0.5 cm
\end{figure}

The difference between critical temperatures for A$_1$ and A$_2$ phases increases with the magnetization. This effect is visible in Fig.\ 4. Increasing Coulomb interaction $U$ creates a growing exchange field in the H-F approximation, which induces increasing magnetization. This behavior was already reported by us \cite{Mizia1}. Spa{\l}ek and co-workers \cite{Wrobel1} have shown a similar effect of change in the superconducting gaps for A$_1$ and A$_2$ phases under an applied field.

\begin{figure}[t]
\begin{center}
\epsfig{file=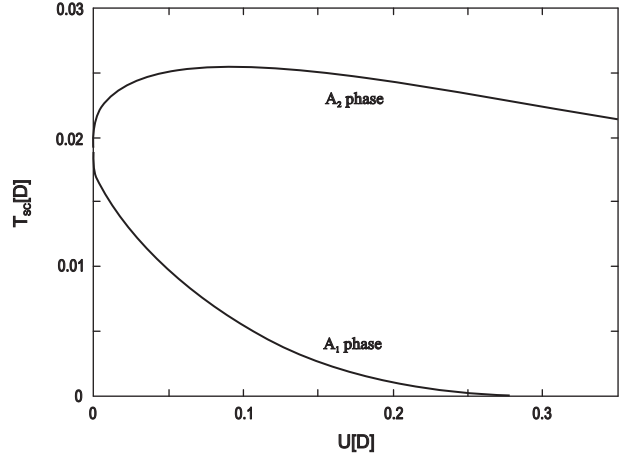,width=0.45\hsize}
	\end{center}
	\caption{Dependence of the superconducting critical temperature $T_{SC}$ for A$_1$ and A$_2$ phase on the Coulomb interaction $U$ in the case of ESP for $n=1.23$, $J=0.0875D$, $S_1=S_2=0.4$, and $D=0.5\, {\rm eV}$.}
\vskip0.5 cm
\end{figure}

\vskip1.0cm 
\noindent {\Large {\bf 5. Influence of external pressure on superconductivity and ferromagnetism}}

\vskip0.5cm 

In the ferromagnetic ZrZn$_2$ compound the ambient pressure level affects the Curie temperature. This compound has the quasi-linear dependence of both magnetic moment and Curie temperature on pressure. The experimental data are reported in Ref.\ \cite{Pfleiderer2}.

In addition to the quasi-linear dependence there is a rapid drop of the magnetic moment at pressure $p_c=16.5\, {\rm kbar}$, which seems to be the first-order phase transition. This transition may be caused by coupling between the long range itinerant magnetization modes and the weak particle-hole excitations. The coupling causes appearance of the non-analytical terms in free-energy near the phase transition point \cite{Belitz1}.

In the ZrZn$_2$ compound the superconducting phase coexists with ferromagnetism. For small pressures the superconducting critical temperature is about 100 times smaller than the Curie temperature.

To explain the pressure dependence of ferromagnetism and superconductivity we assume that the inter-site constants $t$ in the Hamiltonian (\ref{1}) depend on the external pressure. As a result, the kinetic interactions $\Delta t$ and $t_{ex}$ will depend on pressure. They depend on $t(p)$ through Eq.\ (\ref{2}) and on the hopping inhibiting factors $S_1$ and $S_2$, which are pressure dependent. The on-site Coulomb repulsion $U$ is assumed to be pressure independent.

To find the pressure dependence of the hopping integral we need the results for the effective mass $m^*$ dependence on pressure \cite{Lo1}

\be
\frac{1}{{m^* \left( 0 \right)}}\frac{{\partial m^* \left( p \right)}}{{\partial p}} =  - 0.017 \pm 0.004\,{\rm kbar}^{{\rm  - 1}}, 
\label{57}
\ee
hence

\be
\frac{{m^* \left( p \right)}}{{m_0 }} = 1.03\left( {1 - Ap} \right),
\label{58}
\ee
where $m_0$ is the free electron mass and $A \approx 0.017 \pm 0.004\,{\rm kbar}^{{\rm  - 1}}$.

\noindent
Comparing the dispersion relation in the tight binding approximation at small $k$ with the expression for nearly free electrons: $\varepsilon _k  = {{\hbar ^2 k^2 } \mathord{\left/ {\vphantom {{\hbar ^2 k^2 } {2m^* }}} \right. \kern-\nulldelimiterspace} {2m^* }}$, one can obtain

\be
t = \frac{{\hbar ^2 }}{{2m^* a^2 }} = \frac{{\hbar ^2 }}{{2m_0 a^2 }}\frac{{m_0 }}{{m^* }}.
\label{59}
\ee

Assuming for ZrZn$_2$ that lattice constant \cite{Santi1} $a=7.393 \mathop {\rm A}\limits^\circ $  and inserting Eq.\ (\ref{58}) to Eq.\ (\ref{59}) one obtains the relation \cite{Lo1}

\be
t\left( p \right) = \frac{{t\left( {0\,{\rm kbar}} \right)}}{{1 - Ap}}.
\label{60}
\ee
We assume the following pressure relations for the hopping inhibiting factors $S_1$ and $S_2$

\be
\frac{{S_1 \left( p \right)}}{{S_1 \left( {0\,{\rm  kbar}} \right)}} = \frac{{S_2 \left( p \right)}}{{S_2 \left( {0\,{\rm  kbar}} \right)}} = \frac{1}{{1 - Bp}},
\label{61}
\ee  
with parameter $B$ smaller than parameter $A$.

Pressure dependence of: $t$, $t_1$, $t_2$, and of the kinetic interactions: $\Delta t$ and $t_{ex}$, is presented in Figs.\ 5(a) and 5(b), for parameters $A = 0.013\,{\rm kbar}^{{\rm  - 1}}$ and $B = 0.003\,{\rm kbar}^{{\rm  - 1}}$.

\begin{figure}[t]
\begin{center}
\epsfig{file=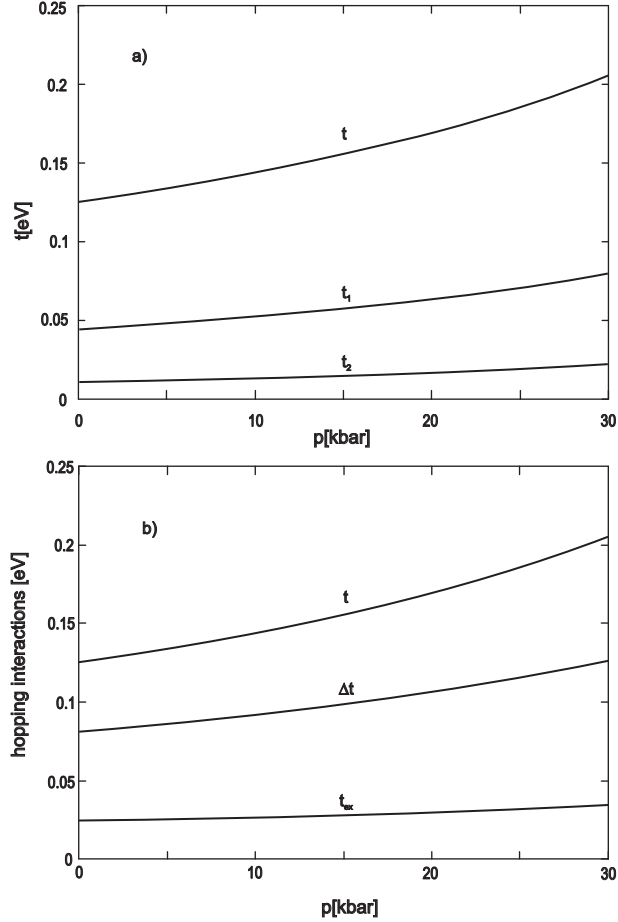,width=0.45\hsize}
	\end{center}
	\caption{Dependence of (a) hopping integrals and (b) kinetic interactions on pressure. The following values were used: $A = 0.013\,{\rm kbar}^{{\rm  - 1}}$, $B = 0.003\,{\rm kbar}^{{\rm  - 1}}$, $S_1 \left( {0\,{\rm kbar}} \right) = 0.35$, $S_2 \left( {0\,{\rm kbar}} \right) = 0.25$, and $t\left( {0\,{\rm kbar}} \right) = 0.125\,{\rm eV}$.}
\vskip0.5 cm
\end{figure}

Figures 6 and 7 show the pressure dependence of the Curie temperature. Figure 6 shows the dependence of $T_C(p)$ for different carrier concentrations. Hopping integral $t(p)$ is expressed by Eq.\ (\ref{60}) and inhibiting factors $S_1$ and $S_2$ are assumed to be pressure independent. The results show a decrease of the Curie temperature with increasing pressure. The Curie temperature always drops to zero with pressure. The exception is at the half-filling concentration. The explanation is following: pressure increase causes an increase of the hopping integral which in turn causes an increase of the critical (minimal) exchange interaction necessary for ferromagnetism while the existing exchange interaction ($U$) remains constant. In Fig.\ 7 we show the dependence of $T_C(p)$ for different values of parameter $B$ in Eq.\ (\ref{61}). At positive values of $B$ the decrease of $T_C$ with pressure is faster than at $B=0$; at $B<0$ we may have even an increase of $T_C$ with pressure.

\begin{figure}[t]
\begin{center}
\epsfig{file=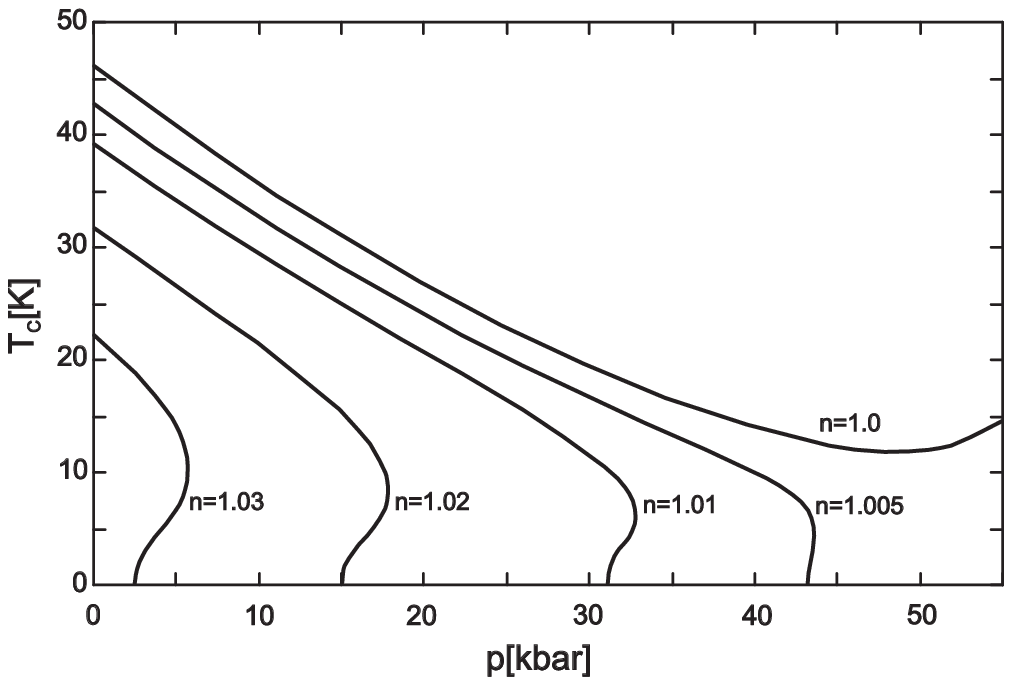,width=0.45\hsize}
	\end{center}
	\caption{Dependence of Curie temperature on pressure for different values of carrier concentrations. The parameters are: $A = 0.017\,{\rm kbar}^{{\rm  - 1}}$, $S_1 = 0.35$, $S_2 = 0.25$, and $t\left( {0\,{\rm kbar}} \right) = 0.125\,{\rm eV}$.}
\vskip0.5 cm
\end{figure}

\begin{figure}[t]
\begin{center}
\epsfig{file=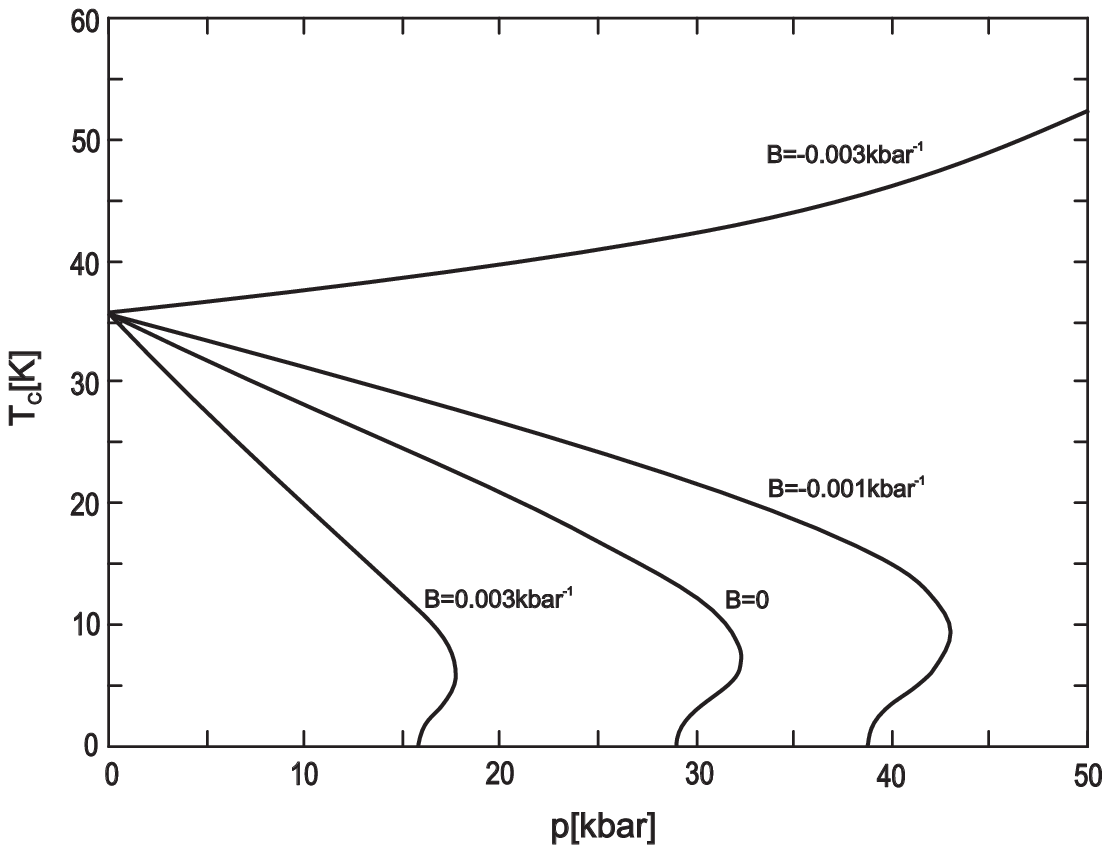,width=0.45\hsize}
	\end{center}
	\caption{Dependence of Curie temperature on pressure for different values of parameter $B$ in Eq.\ (\ref{61}). The other parameters are: $A = 0.013\,{\rm kbar}^{{\rm  - 1}}$, $S_1 = 0.35$, $S_2 = 0.25$, and $t\left( {0\,{\rm kbar}} \right) = 0.125\,{\rm eV}$.}
\vskip0.5 cm
\end{figure}

Papers \cite{Santi1,Walker1,Singh1} show that ZrZn$_2$ has a triplet parallel spin SC. Therefore the calculations will be performed for the coexistence of F with triplet SC. They will be based on Eqs.\ (\ref{53})--(\ref{55}) which give the triplet equal spin superconductivity gap, carrier concentration, and magnetization, respectively. The Fock's parameter will be expressed by the simplified Eq.\ (\ref{49}). At critical temperature we assume $\Delta_k^\sigma =0$ in these equations. In Fig.\ 8 we present the dependence of $T_{SC}(p)$ in the A$_2$ phase at different values of parameter $B$. As shown in the previous Section, the critical temperature in the A$_2$ phase is higher than in the paramagnetic phase. For positive values of $B$ the critical temperature decreases with pressure, while for $B<0$ it increases with pressure.

\begin{figure}[t]
\begin{center}
\epsfig{file=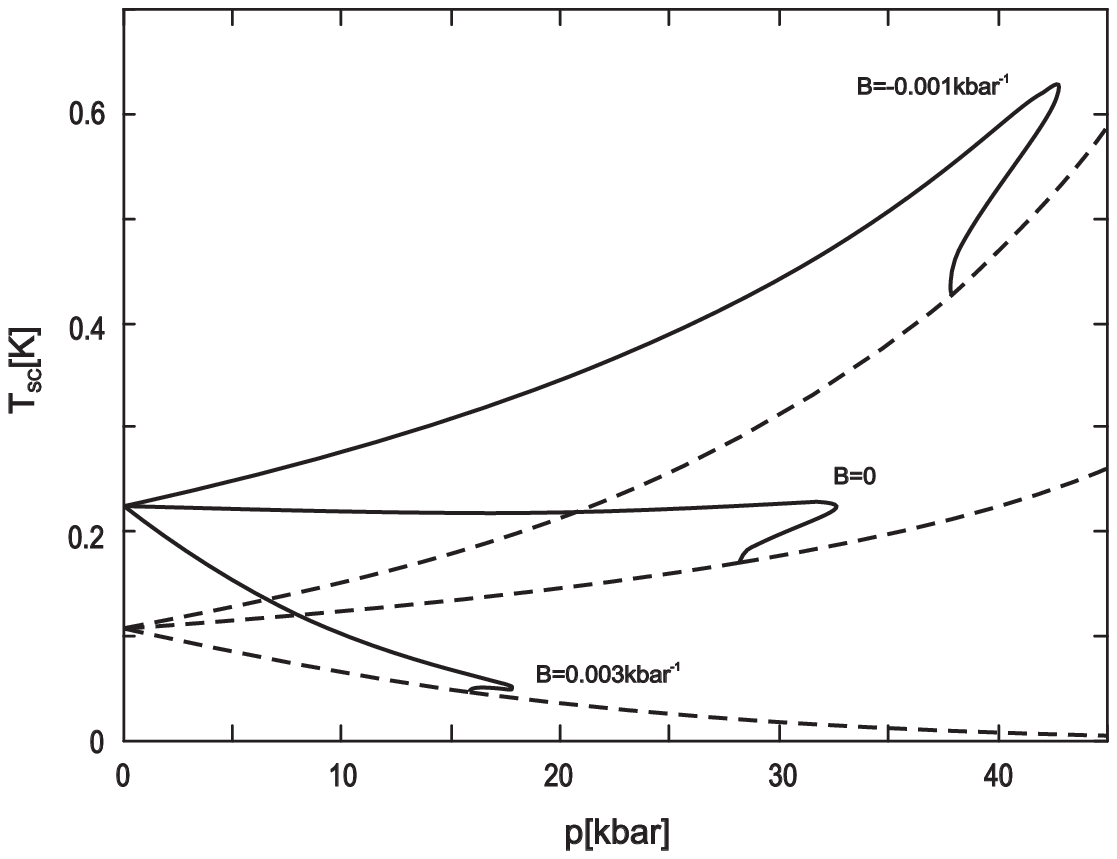,width=0.45\hsize}
	\end{center}
	\caption{Dependence of superconducting critical temperature on pressure for different values of parameter $B$. Solid lines -- $T_{SC}(p)$ at nonzero magnetic moment, dashed lines -- $T_{SC}(p)$ in the paramagnetic state. The other parameters are: $A = 0.013\,{\rm kbar}^{{\rm  - 1}}$, $S_1 = 0.35$, $S_2 = 0.25$, and $t\left( {0\,{\rm kbar}} \right) = 0.125\,{\rm eV}$.}
\vskip0.5 cm
\end{figure}

Figure 9 shows the dependence of Curie ($T_C$) and superconducting ($T_{SC}$) critical temperatures on the pressure. We used the parameters: $A = 0.013\,{\rm kbar}^{{\rm  - 1}}$ and $B = 0.003\,{\rm kbar}^{{\rm  - 1}}$, for which the hopping integrals and kinetic interactions increase with growing pressure [see Figs.\ 5(a) and 5(b)]. On the other hand, the kinetic interactions expressed in the units of bandwidth: $t_{ex}/t$ and $\Delta t/t$, decrease with growing pressure [see Fig.\ 5(b)]. In effect, both critical temperatures for superconductivity and ferromagnetism ($T_{SC}$ and $T_C$) decrease with pressure. This result is compatible with what was observed in ZrZn$_2$.

\begin{figure}[t]
\begin{center}
\epsfig{file=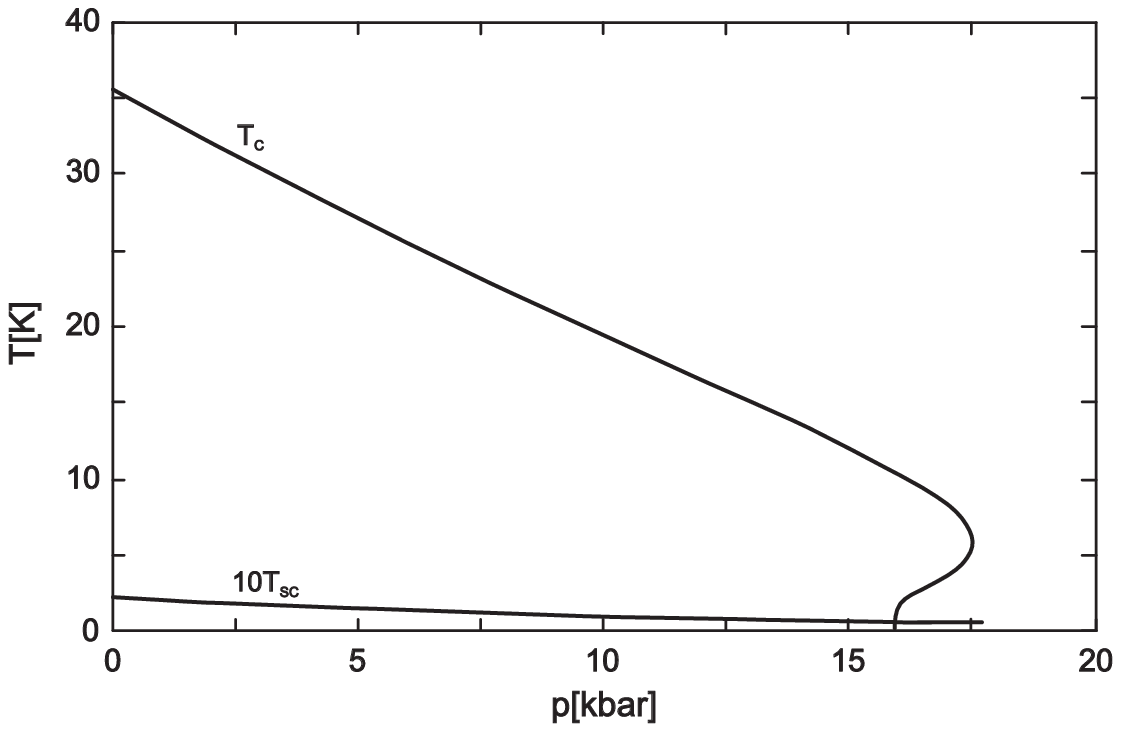,width=0.45\hsize}
	\end{center}
	\caption{Dependence of ESP superconducting critical temperature $10T_{SC}$ and Curie temperature $T_C$ on pressure. The parameters used are: $n=1.015$, $A = 0.013\,{\rm kbar}^{{\rm  - 1}}$, $B = 0.003\,{\rm kbar}^{{\rm  - 1}}$, $S_1 \left( {0\,{\rm kbar}} \right) = 0.35$, $S_2 \left( {0\,{\rm kbar}} \right) = 0.25$, and $t\left( {0\,{\rm kbar}} \right) = 0.125\,{\rm eV}$.}
\vskip0.5 cm
\end{figure}

\vskip1.5cm 
\noindent {\Large {\bf 6. Conclusions}}

\vskip0.5cm 

We have studied the possibility of coexistence between ferromagnetism and superconductivity within the framework of the extended Hubbard model. The main driving forces for these phenomena are the kinetic interactions: hopping $\Delta t$ and exchange hopping $t_{ex}$ interaction. We have analyzed these interactions using the mean-field approximation.

It was found that the hopping interactions influence ferromagnetism in two ways:

-- by increasing the Stoner (exchange) field, which shifts the spin sub-bands with respect to each other. This effect depends on carrier concentration. There is no half-filled band symmetry ($n=1$). The effect is strongest for the smallest carrier concentrations and its strength decreases with increasing carrier concentrations;

-- by modifying the bandwidth factor $b^\sigma$, which depends on magnetization and carrier concentration. Decrease of $b^\sigma$ increases DOS, which favors ferromagnetism. This is a mechanism for raising the ferromagnetism from a gain in kinetic energy rather than from a decrease in potential energy. The effect is in favor of ferromagnetism for carrier concentrations $n>1$.

The kinetic interactions are also driving superconductivity. Singlet s-wave superconductivity is driven by the hopping interaction $\Delta t$, while the triplet superconductivity is driven by the exchange hopping interaction $t_{ex}$.

Analyzing the coexistence of ferromagnetism and the s-wave superconductivity we find that even weak ferromagnetism, if generated by the band shift, destroys the superconductivity. On the other hand the ferromagnetism created by a change of bandwidth can coexist with the singlet superconductivity.

In the case of triplet superconductivity the ferromagnetism creates different critical SC temperatures for the A$_1$ phase (where the pair's spin is parallel to magnetization) and for the A$_2$ phase (where the pair's spin is antiparallel to magnetization). With increasing magnetic moment the difference between both critical temperatures grows. This difference is caused by different bandwidths and different location of the Fermi level in spin sub-bands.

The influence of pressure on magnetism and superconductivity was analyzed based on the assumption that we have: $t(p)$, $S_1(p)$, and $S_2(p)$. An increase of kinetic interactions induced by pressure causes a decrease of the Curie temperature and also a change of critical temperature for triplet superconductivity.

\end{document}